# Deep learning architecture "LightOCT" for diagnostic decision support using optical coherence tomography images of biological samples

A**NKIT** B**UTOLA**[1], D**ILIP** K. P**RASAD**[2], A**ZEEM** A**HMAD**[3], V**ISHESH** D**UBEY**[1], D**ARAKHSHAN** Q**AISER**[4], A**NURAG** S**RIVASTAVA**[4], P**ARAMASIVAM** S**ENTHILKUMARAN**[5], B**ALPREET** S**INGH** A**HLUWALIA**[3, *], **AND** D**ALIP** S**INGH** M**EHTA**[1]

[1]*Bio-photonics Laboratory, Department of Physics, Indian Institute of Technology Delhi, Hauz-Khas, New Delhi- 110016, India.*
[2] *School of Computer Science & Engineering, Nanyang Technological University, Singapore 639798, Singapore.*
[3] *Department of Physics and Technology, UiT The Arctic University of Norway, Norway.*
[4] *Department of Surgical Disciplines, All India Institute of Medical Science, Ansari Nagar, New Delhi-110029, India.*
[5] *Department of Physics, Indian Institute of Technology Delhi, Hauz-Khas, New Delhi- 110016, India.*
*\*balpreet.singh.ahluwalia@uit.no*

**Abstract:** Optical coherence tomography (OCT) is being increasingly adopted as a label-free and non-invasive technique for biomedical applications such as cancer and ocular disease diagnosis. Diagnostic information for these tissues is manifest in textural and geometric features of the OCT images, which are used by human expertise to interpret and triage. However, it suffers delays due to the long process of the conventional diagnostic procedure and shortage of human expertise. Here, a custom deep learning architecture, LightOCT, is proposed for the classification of OCT images into diagnostically relevant classes. LightOCT is a convolutional neural network with only two convolutional layers and a fully connected layer, but it is shown to provide excellent training and test results for diverse OCT image datasets. We show that LightOCT provides 98.9% accuracy in classifying 44 normal and 44 malignant (invasive ductal carcinoma) breast tissue volumetric OCT images. Also, >96% accuracy in classifying public datasets of ocular OCT images as normal, age-related macular degeneration and diabetic macular edema. Additionally, we show ~96% test accuracy for classifying retinal images as belonging to choroidal neovascularization, diabetic macular edema, drusen, and normal samples on a large public dataset of more than 100,000 images. The performance of the architecture is compared with transfer learning based deep neural networks. Through this, we show that LightOCT can provide significant diagnostic support for a variety of OCT images with sufficient training and minimal hyper-parameter tuning. The trained LightOCT networks for the three-classification problem will be released online to support transfer learning on other datasets.

## 1. Introduction

Optical coherence tomography (OCT) is emerging as an increasingly popular technique, which is capable of capturing microscopic and real-time imaging of tissues without exogenous contrast agents. OCT is a non-contact, non-invasive, micron resolution cross-sectional imaging technique, which is proving its potential in various industrial[1, 2] and, biological applications such as ocular disease diagnosis[3], oral cancer[4], breast cancer[5] ovarian cancer[6] and human brain cancer[7], assessment of dental cavities[8], both in *ex-vivo* and *in-vivo*[9, 10]. Because of its high resolution, pathological features can be identified during resection surgery[3]. Spectral-domain optical coherence tomography (SD-OCT) has been used earlier to fit with a needle and allows fine needle-guided biopsy and surgical intervention[11]. Additionally, OCT and Raman spectroscopy was combined to visualize both morphological and biochemical features for tissue characterization[12, 13].

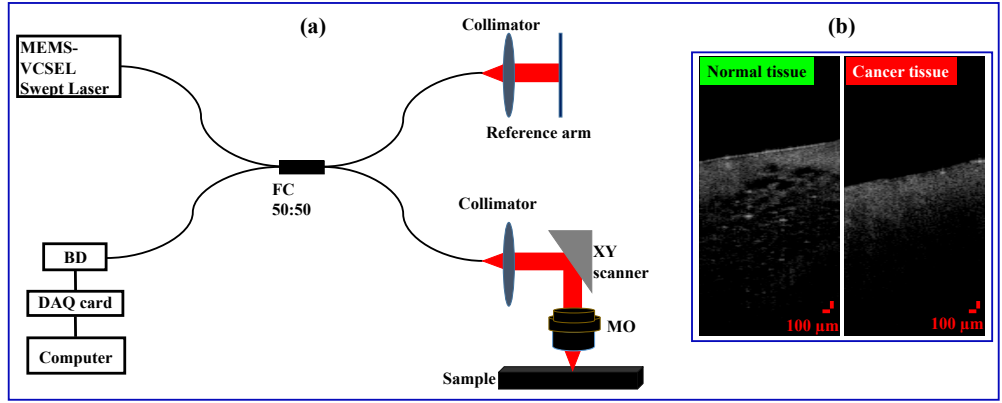

Figure 1: Micro electro mechanical system-vertical cavity surface emitting laser (MEMS-VCSEL) based swept source optical coherence tomography (SS-OCT) (OCS1310V1 - 1300 nm, Thorlabs) system used for imaging normal and cancer (invasive ductal carcinoma (IDC)) breast tissues (a) in the AIIMS dataset. Two illustrative examples of B-scan image (XZ) of normal and IDC breast tissues (b).

Despite various applications of OCT, current diagnostic practice required human expertise to interpret the sample structure and to heuristically derive conclusion i.e., separate them into clinically relevant classes. The diagnostic information is encoded in various forms in the OCT images, and attempts to computationally analyze and provide decision support have been made. Recently, attenuation coefficient map of tissues in the OCT images used for the identification of brain cancer[14]. Textural changes introduced by internal morphology modification during oncogenesis have also been indicated[15]. Volumetric analysis of normal and cancer breast tissues has been done using support vector machine based texture feature analysis[16]. Choi *et al.* have shown full-field optical coherence microscopy applications in quantitative measurement of refractive index distribution for the identification of live cancer cells [17]. Further, certain geometric features in the OCT images, indicative of disease specific morphology, have also been identified as diagnostic indicators[3, 18, 19]. Yet, the diversity of diagnostic features, variations in an imaging system, associated calibrations, and, most importantly, difficulty in deriving a consistent and reliable feature base pose difficulty in applying conventional machine learning and pattern recognition techniques.

Modern deep learning techniques, such as convolutional neural networks (CNN), inherently support abstract characterization of all the physical features discussed above, i.e. texture, refractive index profiles, scattering and absorption coefficient distribution, geometric features, as well as statistics relevant to these features. The ability of CNN has been demonstrated to determine diverse abstract features for classifying a wide variety of objects. CNN is also being adopted for classification of various biomedical images as well[20-22]. A popular approach is to perform transfer learning[23], i.e., use a pretrained CNN (AlexNet, VGG-16, Inception etc.) on object classification datasets and retrain the weights of the CNN for a biomedical dataset[24].

While the approach is simple, it is notable that the weights of pre-trained CNNs are optimized for the classification of objects with crisp object boundaries. Thus, the abstract features encoded in them may not be directly relevant or optimal for biomedical images, which often sport fuzzy regions and spread out diagnostic features. Consider, for example, the swept-source OCT (SS-OCT, see Fig. 1a) (OCS1310V1 - 1300 nm, Thorlabs) images of normal and cancer (invasive ductal carcinoma (IDC)) breast tissues taken at All India Institute of Medical

Science (AIIMS) Delhi and shown in Fig. 1b. AIIMS dataset captured for breast tissue classification using custom deep learning architecture i.e., LightOCT. Details about the AIIMS dataset are given in the section 2.2. Cross-verification of the data is performed in AIIMS using histopathology. From Fig. 1b, it is difficult to mark the IDC region as compared to the healthy tissue even difficult to identify relevant texture features that can differentiate normal and cancerous tissues using simple machine learning techniques. Moreover, the complete lower half of the images corresponds to tissues, which makes it difficult to localize the abstract pre-learnt features that characterize the two tissue classes. In simple terms, the features that allow differentiation of these tissues are distributed across the entire image.

Here, we propose "LightOCT" architecture, customized for OCT image classification. LightOCT is a simple architecture that can classify various OCT datasets with very few tunable hyperparameters. The simplicity of LightOCT with a few hyper-parameters allows easy customization for individual datasets. Second, it provides insight into the texture kernels which have a consequence for classification, for interpretation by pathologists. Third, we show applicability of LightOCT for three diverse and independent OCT datasets:

(i) AIIMS dataset of IDC and normal breast tissues.
(ii) Srinivasan's dataset of OCT images of aged, diabetic, and normal retinal tissues[18].
(iii) Kermany's dataset of OCT images of choroidal neovascularization, diabetic macular edema, drusen, and normal tissues[3, 25].

Through this work, we show that LightOCT can provide significantly high accuracy for cancer detection and classification between different types of retinal diseases. The training time and performance of the LightOCT architecture is also compared with three different networks i.e. VGG-19, ResNet-101 and Inception-V3. The current approach will be an important step towards removing the barrier between artificial intelligence and clinical applications. Also, LightOCT can be implemented directly in OCT imaging for diagnosis, risk stratification, and prognosis of several diseases. The results demonstrate it as a valuable tool for decision support system.

## 2. Methods

### 2.1. Datasets

#### 2.1.1. AIIMS dataset

AIIMS dataset is a dataset of SS-OCT (OCS1310V1 - 1300 nm, Thorlabs) images of normal and cancerous breast tissues. The dataset initially released with a volumetric analysis of breast cancer tissue using support vector machine based texture feature analysis[16]. The datasets were collected from 22 patients at All India Institute of Medical Science (AIIMS) New Delhi. All diseased samples were histologically confirmed cases of invasive breast cancer. These patients operated to remove the tumor. The types of surgery were mastectomy and breast conservation, from where the tumors removed. The diseased samples were prepared from the core of the tumor, whereas normal tissue samples prepared from the normal area of the breast far beyond the tumor margin. Informed consent obtained from all the patients before the experiment. The imaging protocol for human tissue studies was approved by the ethical committee of the Indian Institute of Technology (IIT) Delhi and AIIMS Ethics committee. Ethics committee of AIIMS functions as per ICH GCP and other applications regulatory guidelines. After preparing the sample both for normal and cancerous tissues, the remaining part of the tumors was sent for histopathology. A senior consultant pathologist later confirmed the histology of the normal and cancer samples.

Further, to record the OCT dataset, two volumetric OCT images were acquired from each diseased and normal tissue sample at two different locations. Each volumetric OCT image correspond to 2 mm × 2 mm × 2.5 mm in *x-y-z*-direction and contains 105 B-scan (2D) images.

Each B-scan OCT image represents a different XZ plane and not a cross-sectional (*en-face*) image. Some of the B-scan OCT images were discarded to avoid measurement inaccuracy of the system. These scans suffers with unwanted distortion and artifact due to mechanical inertia in the galvanometer-based scanner [26]. Additionally, each B-scan image is taken as a distinct image to perform the classification study.

MEMS VCSEL SS-OCT imaging system was used for imaging normal and cancerous breast tissue. Schematic diagram of MEMS VCSEL SS-OCT (OCS1310V1 - 1300 nm, Thorlabs) system consisting of a light source module (laser), an imaging module, and a standalone probe can be seen in Fig. 1 (a). Swept source of central wavelength 1300 nm was used to perform the study. Source contained bandwidth of 100 nm includes Mach-Zehnder interferometer "k-clock" for optical clocking data acquisition[27]. A 5X objective lens (MO, LSM03, Thorlabs, focal length ~ 25.1 mm in the air) was used in the sample arm to image the specimen under study. Dual-balanced photodiode at 100 kHz A-Scan rate was used to detect interference signal. Two-dimensional XY scanner is used to scan the probe beam over the sample to get the 3D image. Axial and transverse resolution of the system is 12 μm and 16 μm in air, respectively. Laser with an average output power of 25 mW is used to illuminate the sample.

### 2.1.2. Srinivasan's dataset

Srinivasan's dataset[18] contained 45 samples (15 normal, 15 dry age-related macular degeneration (AMD) and 15 with diabetic macular edema (DME)) captured by spectral domain-OCT (SD-OCT) system (Heidelberg Engineering Inc., Heidelberg, Germany). The SD-OCT system, which is used to acquire the volumetric image of all these samples, offers 3.87 μm axial resolution while the lateral resolution is varying from 6-12 μm. Scan dimensions for the datasets varying from $5.8 \times 5.8$ mm$^2$ to $9.1 \times 7.6$ mm$^2$. The number of A-scan and B-scan varying according to the scanning dimensions of the study objects, while each image is cropped to the center 150 column (pixels) in a lateral direction and 45 pixels in an axial direction to perform the study. All datasets are acquired in Duke University, Harvard University, and the University of Michigan. We have used the dataset as it is provided by the source and not manipulated the images before using them for training or testing in this work.

### 2.1.3. Zhang's dataset

In Zhang's datasets, spectral-domain OCT (SD-OCT) used to capture *in vivo* three-dimensional images of retinal tissue. The datasets contained 109,312 OCT images from 5,319 patients (no criteria of age and gender). A total of 37,456 images of the datasets belong to choroidal neovascularization, 11,599 with diabetic macular edema, 8,867 with drusen and 51,390 normal. The OCT datasets are acquired from five different hospitals and eye center (the Shanghai First People's Hospital, Eye Institute of the University of California San Diego, Eye Institute of the University of California San Diego, Beijing Tongren Eye Center and Medical Center Ophthalmology Associates) between July 1, 2013 and March 1, 2017.

### 2.2. Data analysis

LightOCT is implemented in Matlab on a 64-bit Windows OS. The machine configuration is Intel Xeon CPU E5-1650 v4 @ 3.6 GHz with 128 GB RAM and Nvidia 1080 Ti GPU. For AIIMS datasets, different patient's volumetric OCT images are used for training and testing purpose. As mentioned in section 2.1.1 that 2 volumetric OCT images were acquired from each diseased and normal tissue sample at two different locations. Each volumetric OCT image contains 105 B-scan (2D) images. The volumetric datasets of 40 normal and 40 cancer OCT images are used for training with 10-fold cross-validation, and 4 normal and 4 cancer OCT images for testing purpose. Cross-validation is a resampling procedure used to evaluate any machine learning model in general. 10-fold cross validations imply that the training data partitioned into "10" random subset. One subset is used to validate the model trained using the remaining subsets. After the validation, the trained model is used for testing purpose. Testing of the network is done on 4 normal and 4 cancer volumetric OCT images, different from the data used in the training datasets. To rule out the impact of choice of training, validation, and test datasets, we independently run the whole process of a train, validate and test using randomized patient selection for 20 times.

Classification results for Srinivasan's datasets are obtained by randomly assigning ~73% datasets i.e. 33 patient's OCT images (11 Normal, 11 DME and 11 age related macular degeneration) for training and 12 patient's OCT images (4 Normal, 4 DME and 4 age related macular degeneration) for testing purpose. Further, standard data augmentation techniques is applied to perform the classification study. Data augmentation is a preprocessing step for image augmentation such as rotation, resizing and reflection in Deep Learning Toolbox$^{TM}$ of Matlab 2019b. In our case, we use random reflection, range of uniform scaling and range of vertical and horizontal translation to augment the datasets. For Zhang's datasets, 108,312 OCT images (27,206 CNV, 11,349 DME, 8,617 Drusen and 51,140 normal) from 4,686 patients are used to train the network and 1,000 OCT images (250 CNV, 250 DME, 250 Drusen and 250 normal) from 633 patients are used for testing purpose. Stochastic gradient descent with momentum (SGDM) is used for training the CNN. The initial learning rate is set as 0.0001 and is kept adaptable in the process of learning using default settings of the SGDM learning code of Matlab. The maximum number of epochs in the learning process is set as 20. The Default value is assigned to the other parameters of the learning process. Training of the network is done from scratch for each dataset. We recommend doing that for each different instrument, and each different biological problem since the characteristic features may vary greatly, and the pre-trained network may be biased to previously encoded features. The percent values reported in the manuscript corresponds to the recall (also called sensitivity) for each class, i.e., a ratio of the number of true positives classifications for a class to the number of images with this class label in the ground truth.

### 2.3. LightOCT

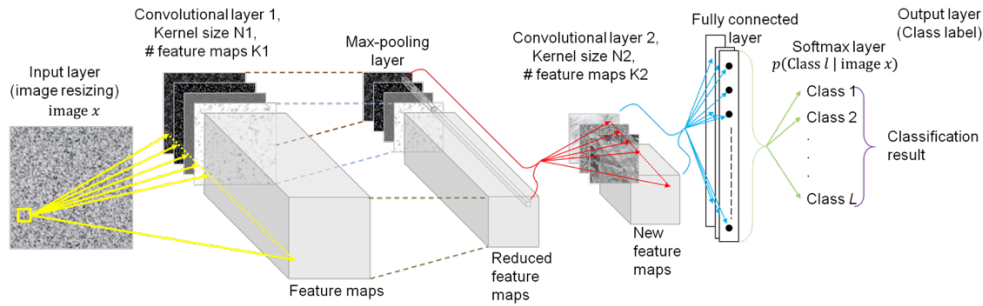

Figure 2: LightOCT architecture for the classification of different OCT images. LightOCT is used for classification of AIIMS dataset (normal and cancer breast tissue), Srinivasan dataset (normal, age-related macular degeneration and diabetic macular edema) and Zhang dataset (choroidal neovascularization, diabetic macular edema, drusen and normal samples). It features two convolutional layers and one fully connected layer. The hyper-parameters, N1, K1, N2 and K2 of the convolutional layers of LightOCT are tunable where N shows the size of kernel and K represents number of feature maps.

LightOCT is a CNN with 2 sets of convolutional layers and a fully connected layer, followed by a soft-max layer for classification. Its architecture is shown in Fig. 2, and the details of the layers are given in Table 1. The input size of the network as shown in Table 1, decided empirically to achieve better performance of the network for all three datasets. It has been shown in the previous study that input size certainly affect the classification accuracy of the network [28]. The first convolutional layer identifies local texture features in the vicinity of each pixel. Each neuron in this layer corresponds to one spatial window centered at a pixel in the input image and a kernel in this layer. Activation of each neuron in this layer corresponds to the presence of the texture features, represented by the corresponding kernels. The rectifier linear unit (ReLU) then clips the output so that the network remains stable. Thus, for each kernel, we get one feature map, which is approximately the same size as the input image. The feature map is reduced to spatially half the size in the next layer, which is the maxpooling layer. It simply retains the feature value, which is the maximum in a region of 2 x 2 pixels. This implies that the maximum activation in each local region of 2 x 2 pixels are retained for further analysis. The next layer is a second convolutional layer. This layer identifies the local spatial features in the texture feature maps generated by the previous layers. In essence, it looks for local spatial patterns pertaining to the presence of the texture features using the kernels in the second layer. The fully connected layer after this layer combines the activations of all the neurons in the previous layer. The number of fully connected layers varies as a number of classes in the datasets i.e. 2, 3 and 4 in AIIMS, Srinivasan and Zhang datasets, respectively. It computes cross-spatial and cross-feature relationships to generate activations for each class. In fully connected layer, the information of even distantly located features are combined and assessed for a given class. This layer is followed by a softmax layer, that computes the relative activations of all the classes using the softmax function on the activations of the fully connected layer. Lastly, the class label is identified in the output layer as the class for which the softmax function generates the maximum activation. Thus, the main functional architecture of LightOCT can be described as the first convolutional layer computing the texture features using textures represented in its kernels, second convolutional layer computing the local cross-feature patterns, and the fully connected layer employing cross-spatial and cross-feature patterns to result into differentiability of the classes.

Table 1. The details of the architecture of LightOCT are given below, including the parameters of LightOCT.

| Layer name | Function | Number of weights | Output data size |
| --- | --- | --- | --- |
| Input layer | Resizes images internally to size 245 x 442 pixels | Nil | 245 x 442 |
| Convolutional layer 1 (plus rectified linear unit (reLu) and not illustrated in Figs. 4 for brevity) | Apply K1 convolution kernels of size N1 x N1 on each pixel of the input image (with stride 1), followed by rectification. | N1 x N1 x K1 (N1 = 5, K1 = 8) | 243 x 440 x8 |
| Max pooling | Down sample the output of previous layer. | Pool size 2 | 121 x 220 x 8 |
| Convolutional layer 2 (plus ReLU) | Apply K2 convolution kernels of size N2 x N2 on each pixel of the input image (with stride 1), followed by rectification. | N2 x N2 x K2 (N2 = 5, K2 = 32) | 119 x 218 x 32 |
| Fully connected layer | Combine cross-feature and cross-space information to activate one neuron corresponding to each class. | 830144 x L (830144 corresponds to all pixels in output of previous layer, L corresponds to number of classes) | L |
| Softmax layer | Computes the softmax function of the output of the previous layer for each class. | | L |

| Output layer | Computes which class has the maximum value at the output of the softmax layer. | | 1 (class label) |

## 3. Results

### 3.1 Functioning of LightOCT

The functioning of LightOCT is illustrated in Fig. 3 using the example images shown in Fig. 1(b). The values of activations of neurons in fully connected layers (for each class) and the conversion to relative scores for making final decisions are explicitly shown for the two examples. Figure 3 shows the abstract nature of the hidden layers. The feature maps derived from convolutional layers do not visually indicate that the distinguishability of the normal and cancer tissue images. However, the net effect of texture features (the first convolutional layer), local-cross feature patterns (the second convolutional layer), and cross-spatial cross-feature patterns (the fully connected layer) is evident in the outputs of the fully connected layer. The fully connected layer's outputs for the two class labels do have a large difference between them for each image, but the relative strength of the conclusion regarding each class label is not apparent. This conversion from the abstract output of a fully connected layer to a human interpretable conclusion is derived through the softmax layer.

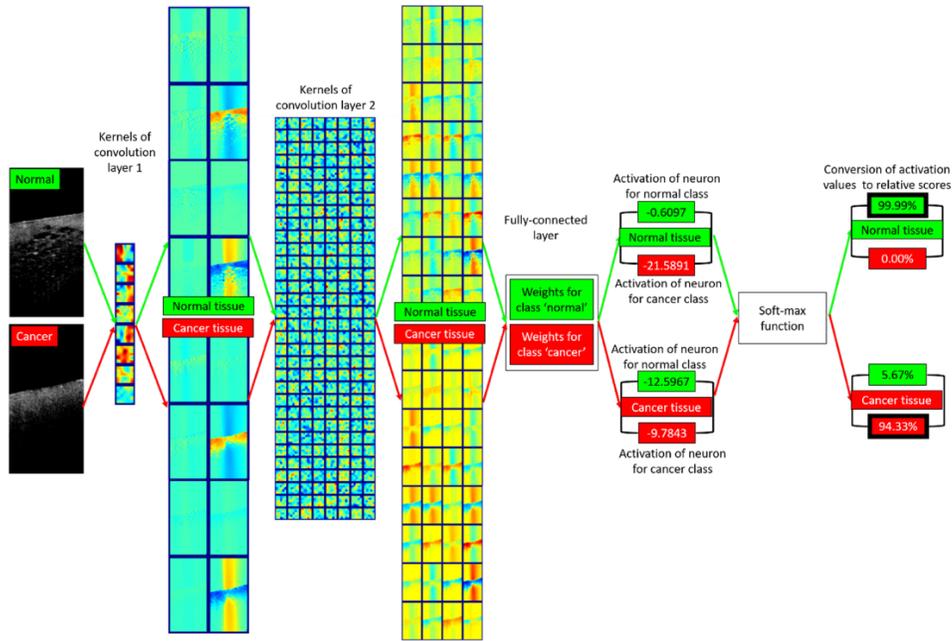

Figure 3: Illustration of the critical steps of LightOCT for normal and IDC cancer breast tissue OCT image of AIIMS dataset is shown here. Total 8 and 32 kernels are shown in the first and second convolution layer, respectively. The net effect of texture features (the first convolutional layer), local-cross feature patterns (the second convolutional layer), and cross-spatial cross-feature patterns (the fully connected layer) is evident in the outputs of the fully connected layer. The result of rectifier linear units and intermediate layers are not shown here for simplicity.

### 3.2 Hyper-parameters of LightOCT

LightOCT has four hyper-parameters $N_1$, $N_2$, $K_1$, and $K_2$. In general, we choose $N_1 = N_2$. We first show the effect of $N_1$ on the classification performance on all the three datasets. The classification results (recall in %) for three values of N1 are given in Table 2. It can be seen that N1, N2 = 5 gives the best result across all the datasets. We expect that the reason is that a kernel of size 5 is just the right size for representing textures in these datasets. We first discuss this result from the perspective of decision support. Since these datasets are from independent sources, it might indicate that the texture features in OCT images for these biological structures, i.e., breast tissue and ocular tissue, are of these scales. Unsurprisingly, thus, explicit identification of such small texture features visually by even human experts is difficult.

Table 2. Classification results (recall values in %) for the AIIMS dataset (Normal and IDC breast tissue), Srinivasan datasets (aged, diabetic and normal retinal tissues) and Zhang datasets (choroidal neovascularization, diabetic macular edema, drusen and normal). N1 and N2 represents the size of kernel in first and second convolutional layer of the network, respectively.

|  | N1, N2 = 3 | N1, N2 = 5 | N1, N2 = 7 |
|---|---|---|---|
| **AIIMS dataset** | | | |
| Normal breast tissue | 91.2% | 99.3% | 89.3% |
| Cancerous breast tissue | 93.5% | 98.6% | 99.8% |
| **Srinivasan dataset** | | | |
| Aged retinal tissue | 97.3% | 98.4% | 98.4% |
| Diabetic retinal tissue | 98.4% | 99.2% | 98.0% |
| Normal retinal tissue | 98.2% | 98.7% | 98.2% |
| **Zhang dataset** | | | |
| Choroidal neovascularization | 86.4% | 96.8% | 89.2% |
| Diabetic macular edema | 73.6% | 93.2% | 64.4% |
| Drusen | 46.8% | 90.4% | 57.6% |
| Normal | 91.6% | 97.6% | 92.4% |

To understand the results from machine learning aspect, kernels for different values of N1 for first convolution layer for the AIIMS dataset are shown in Fig. 4. As evident, kernels for N1 = 3 do not represent textural features of sufficient variety over space. The kernels for N1 equal to 5 or 7 show more spatial variety. However, kernels for N1 = 7 show large variations within each kernel. This indirectly means that 8 feature kernels (K1 = 8) may not be sufficient in representing the required diversity of features for a kernel size of 49 pixels. The number of weights to be learnt for a kernel size of N1 and feature size of K1 for a single channel image is $N1^2 K1$. Thus, from the perspective of dimensionality of learning also, it is preferable to choose N1 = 5.

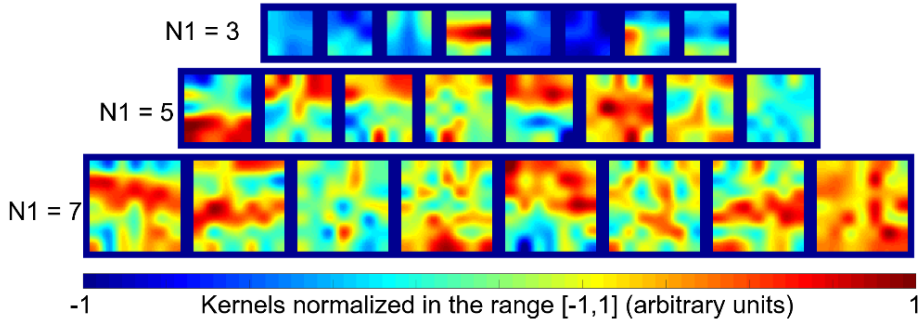

Figure 4: Kernels learnt by LightOCT for three different values of N1 are shown here for the AIIMS dataset. The kernels are resized 5 times with smoothing for the ease of visualization. These represent the texture features identified by the first convolutional layer.

The effect of the hyper-parameters K1 and K2 is shown in Table 3. The parameter K1 indicates the number of texture features learnt directly from an OCT image. Thus, increasing the value of K1 implies that more texture features can be learnt, which can translate to better accuracy. At the same time, the number of weights in the first convolutional layer is $N1^2 K1$. Thus, increasing K1 implies that large number of weights must be learnt. This increases the computational load of learning, as well as the chances of underfitting or mis-convergence. In other words, a compromise is sought in the accuracy and computational load. Similar considerations apply for the parameter K2 as well. Table 3 shows the comparison between classification accuracy and the number of weights learnt for the AIIMS dataset with different values of K1 and K2. We found that (K1 = 8, K2 = 32) is sufficient for all the datasets. Especially, the accuracy for the more complex 4-class problem, such as Zhang's dataset, is better for the combination K1 = 8, K2 = 32. This reveals the importance of choosing just the sufficient number of features and not resorting to standard CNNs pre-learnt on data acquired from significantly different instrument. Thus, we expect that the LightOCT is better suited for OCT image datasets and easily customizable for reliable performance on the dataset derived from an instrument for specific medical conditions in the given racial diversity of a particular region. Further, optimizing different hyperparameters such as number of kernels, size of kernels, input image size, batch size etc. in the customized architectures affect the classification accuracy of the model. However, these parameters cannot be fine-tuned effectively in state-of-the-art model and also required heavy computational load and training time to classify the images.

Table 3. The classification results (recall in %) of LightOCT for the AIIMS dataset (Normal and IDC breast tissue), Srinivasan datasets (aged, diabetic and normal retinal tissues) and Zhang datasets (choroidal neovascularization, diabetic macular edema, drusen and normal tissue) using different values of K1 and K2 is shown here.

|  | K1 (N1 = 5, K2 = 32) | | K2 (N1 = 5, K1 = 8) | |
|---|---|---|---|---|
|  | 8 | 16 | 32 | 64 |
| **AIIMS dataset** | | | | |
| Normal breast tissue | 99.3% | 99.0% | 99.3% | 90.7% |
| Cancerous breast tissue | 98.6% | 88.1% | 98.6% | 94.5% |
| **Srinivasan's dataset** | | | | |
| Aged retinal tissue | 98.4% | 95.6% | 98.4% | 97.8% |
| Diabetic retinal tissue | 99.2% | 97.5% | 99.2% | 98.4% |
| Normal retinal tissue | 98.7% | 99.2% | 98.7% | 99.5% |
| **Zhang dataset** | | | | |
| Choroidal neovascularization | 96.8% | 90.4% | 96.8% | 87.6% |
| Diabetic macular edema | 93.2% | 71.2% | 93.2% | 73.6% |
| Drusen | 90.4% | 56.4% | 90.4% | 61.6% |
| Normal | 97.6% | 90.8% | 97.6% | 93.2% |

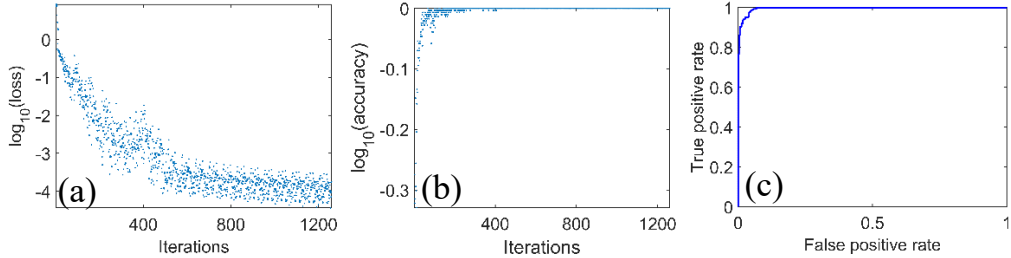

Figure 5: The training and performance characteristics of LightOCT for normal and cancer breast tissue is shown here: (a) Loss curve, (b) accuracy curve and (c) receiver operating characteristics (ROC) curve.

The training and performance characteristics of LightOCT for the AIIMS dataset is shown in Fig. 5. The training loss and training accuracy help us to understand the learning of the network after each iteration. The receiver operating characteristic (ROC) curve depicts in Fig. 5(c). ROC curve is an important evaluation for checking any classification model performance. The ROC curve is plotted between false positive rate (1-specificity) in x axis and true positive rate (sensitivity) in y axis. The other performance measures, including training and test time are reported in Table 4. The area under curve (AUC) for the ROC curve shows the testing performance of the model. The training time for the AIIMS datasets was 1025 sec which is very less compare to the transfer learning approach. The LightOCT architecture take less training time because of only 2 convolution layers presented in the network. The performance parameter shown in Table 4 shows the fast and robustness of the architecture.

Table 4: Performance parameters of LightOCT for the AIIMS dataset with 'cancer' as the posterior class.

| Recall (sensitivity) | 98.6% |
|---|---|
| Precision | 99.3% |
| Specificity | 99.1% |
| False omission rate | 2.1% |
| Overall accuracy | 98.8% |
| F-score | 98.9% |
| Area under curve (AUC) for the ROC curve | 99.6% |
| Time taken for training (GPU) | 1025 s |
| Time taken for training in CPU (no GPU) | 4719 s |
| Test time per image (batch process) in CPU (no GPU) | 6 ms |
| Test time per image (single image process) in CPU (no GPU) | 94 ms |

## 3.3 Comparison between LightOCT and transfer learning

The comparison between LightOCT and transfer learning for classification of three different OCT datasets is shown in Table 5. VGG-19, ResNet-101 and Inception-V3 architecture implemented for transfer learning purpose. Similar training and testing procedure followed for VGG-19, ResNet-101 and Inception-V3 as for LightOCT network. Table 5 shows that LightOCT, VGG-19, ResNet-101 and Inception-V3 provide and overall classification accuracy of 98.5%, 96.3%, 93.9% and 93.5%, respectively for AIIMS dataset. Similarly, for Srinivasan's datasets, the performance of VGG-19 and ResNet-101 is found comparable and Inception-V3 offers 1% higher accuracy (99.9%) than the LightOCT (98.8%) network. Finally, the overall accuracy of 94.5%, 93.3%, 96.7% and 95.8% is achieved for Zhang datasets by LightOCT, VGG-19, ResNet-101 and Inception-V3, respectively. For Zhang datasets, ResNet-101 provide the best classification accuracy which is 2% higher than the proposed LightOCT network.

Table 5: Comparison between LightOCT and transfer learning for classification (recall in %) of three different OCT datasets.

|  | LightOCT | VGG-19 | ResNet-101 | Inception-V3 |
|---|---|---|---|---|
| **AIIMS dataset** | | | | |
| Normal breast tissue | 99.3% | 98.8% | 100% | 100% |
| Cancerous breast tissue | 98.6% | 93.8% | 87.8% | 87.1% |
| **Srinivasan dataset** | | | | |
| Aged retinal tissue | 98.4% | 98.4% | 100% | 100% |
| Diabetic retinal tissue | 99.2% | 95.1% | 100% | 100% |
| Normal retinal tissue | 98.7% | 93.0% | 97.9% | 99.7% |
| **Zhang dataset** | | | | |
| Choroidal neovascularization | 96.8% | 97.2% | 98.8% | 99.2% |
| Diabetic macular edema | 93.2% | 90.4% | 100% | 99.2% |
| Drusen | 90.4% | 86.8% | 88.4% | 86.8% |
| Normal | 97.6% | 98.4% | 99.6% | 98.0% |

On the other hand, training time for VGG-19, ResNet-101 and Inception-V3 is found very higher than the LightOCT network. Table 6 shows the comparison of training time for each network for different datasets. For example, in Srinivasan's datasets, ResNet-101 took 4,753 sec for the training, which is almost 13 times higher than the training time of LightOCT network with almost similar performance. For Zhang datasets, the training time for LightOCT, ResNet-101 and Inception-V3 was 18,063 sec, 2,52,033 sec and 1,98,111 sec, respectively. Additionally, despite consisting only 2 convolutional layers, the overall performance of LightOCT is found comparable to the state-of-the-art technique which conclude that complex networks are not necessarily required to achieve better performance in every classification problem.

Table 6: Comparison of training time for LightOCT, VGG-19, ResNet-101 and Inception-V3 for three different OCT datasets.

| Deep neural network (total learnable parameters in millions) | Training time (sec) | | |
|---|---|---|---|
| | AIIMS dataset | Srinivasan dataset | Zhang dataset |
| LightOCT (0.8) | 1,025 | 374 | 18,063 |
| VGG-19 (144) | 5,800 | 1,336 | 74,490 |
| ResNet-101 (45) | 21,811 | 4,753 | 2,52,033 |
| Inception-V3 (24) | 18,754 | 4,409 | 1,98,111 |

### *3.4 Class activation maps for various classes*

Figure 6 depicts the class activation maps (CAM) of each class of OCT image in three different datasets. The CAM indicates the discriminative region used by LightOCT to differentiate between different datasets. In other words, CAM [29] represent the firing strength of various regions that help in the conclusion regarding the class label. For man-made objects with define boundaries in the training sample/test sample, researchers have tried to visualize and understand the functionality of the CNN model via activation maps. Recently, Jing et al.,[30] shows the activation map of OCT images using VGG-16, ResNet-50 and Inception-V3 architectures which contained 138, 26 and 24 million parameters, respectively. These parameters contribute to plot CAM of the image. Further, VGG-16, ResNet-50, and Inception-V3 has 41, 177 and 316 layers, respectively. On the other hand, due to only 2 convolutional layers and total 9 layers architecture, LightOCT contained 0.8 million parameters which are far less than the conventional DNN architectures. It is known that first convolution layer features of a DNN look more like edge features and second convolution layer will look slightly little more than simple edge features. Hence, it may or may not highlight the regions that are obvious to human interpretation. Nonetheless, the key finding of our proposed study shows that fine tuning to 0.8 million parameters is sufficient to differentiate between 2-4 classes of OCT images. Therefore, a deeper network with millions of parameters is not required in every classification problem.

Figure 6 compare CAM of the OCT image using LightOCT, after second convolution layer of Inception-V3 and before the fully connected layer of Inception-V3. CAM is plotted for all three different datasets i.e. AIIMS, Srinivasan and Zhang datasets. The activation map after second convolution layer of Inception-V3 (Fig. 6($a_2$-$i_2$)) and just before the fully connected layer of Inception-V3 (Fig. 6($a_3$-$i_3$)) shows the clear dependence of activation map on the number of deeper layers. Since the activation map in Fig. 6($a_1$-$i_1$) and Fig. 6($a_2$-$i_2$) represents the weighted sum to the parameters in two convolution layers only, it will not be similar as final activation map of Inception-V3. In addition, CAM images of LightOCT are highlighting few regions in AMD, DME, CNV and drusen which might be sufficient for the network to differentiate between normal and diseased OCT image. It might be a possibility that further increasing the number of classes will decrease the accuracy of LightOCT and then there will be a requirement to add more layers in the network.

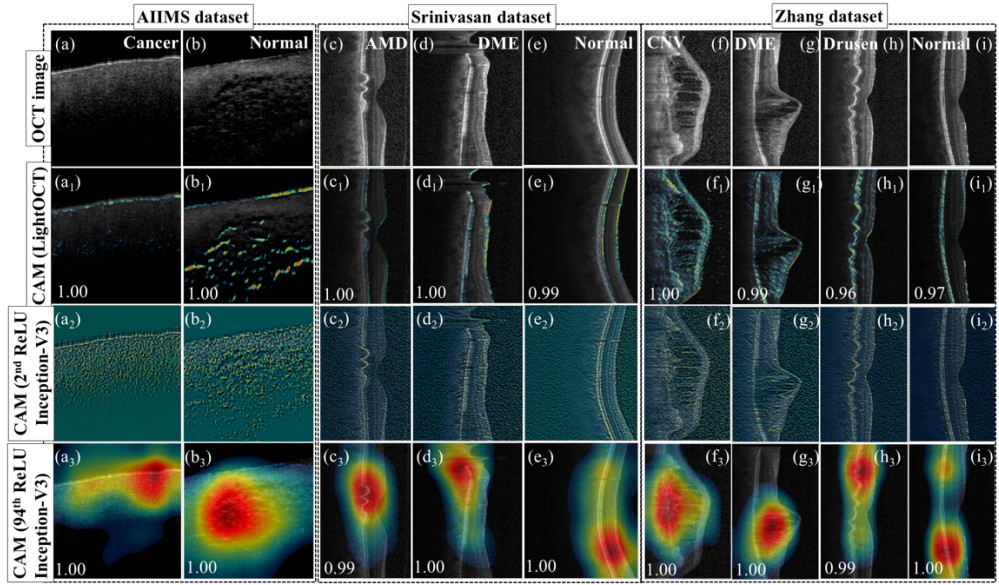

Figure 6: Class activation map and prediction scores of normal/cancerous (AIIMS datasets) and ocular disease (Srinivasan and Zhang datasets) OCT images. (a-i): OCT images of three different datasets. The activation map is shown for LightOCT($a_1$-$i_1$), after second convolutional layer of Inceptions-V3 ($a_2$-$i_2$) and just before the fully connected layer of Inception-V3 ($a_3$-$i_3$) network.

## 4. Discussion and conclusion

We propose LightOCT, a CNN architecture, can provide an excellent accuracy for different OCT image datasets. We show the effect of tuning hyper-parameters of LightOCT on the performance, especially indicating that the number of features in convolutional layers or larger kernel sizes do not necessarily translate to better accuracy. Our results demonstrate that the architecture provides excellent accuracy of classification on three independent datasets, targeting 2 to 4 class classification problem for breast tissues and ocular tissues, targeting diverse conditions of significance in clinical diagnosis. We report 98.9% accuracy for the AIIMS dataset for the classification of normal and cancer tissues. More than 96% accuracy is reported for Srinivasan's dataset in classifying tissues as aged, diabetic, or normal tissues. There is a small decrease in the accuracy, which can be explained by the increased difficulty in classifying 3 classes. LightOCT uses only a single scan at a time and still provides correct classification for more than 96% over more than 3000 individual scans.

The main advantage of LightOCT is to achieve similar performance as complex and advanced deep learning architectures like VGG19, ResNet-101, and inception-V3 for the OCT datasets. Further, due to its lightweight and low training time and low computational need, the usability of LightOCT is higher compared to other advance architectures. LightOCT can be trained on basic laptop CPU in acceptable training time (4,719 sec for AIIMS datasets) while advance DNN will need more computation system and training time will be few days on CPU system. Secondly, our proposed study shows that fine tuning of a smaller number of parameters i.e. 0.8 million in LightOCT is sufficient to differentiate between 2-4 classes of OCT images. Therefore, a deep network with millions of parameters is not required in every classification problem. Finally, the input size of an image certainly affects the accuracy of the network, but the needed computations grow quadratically[28]. In addition, for larger input image, the performance of the network improved mostly because of the larger spatial size of the deeper layer[28].

We anticipate that LightOCT will be a practical architecture that can be trained on datasets at local clinics and integrated as classify-while-you-image model easily using general computational system. Lastly, we opine that custom architectures designed for specific types of microscopes, such as LightOCT for OCT images, is a better approach than simply performing transfer learning on existing architectures pre-trained on unrelated problems. This is in consistence with Xing et. al. [31], who conclude that shallow networks may be better for microscopy images as compared to deep networks. At first sight, it may appear in conflict with Tajbaksh et. al. [32] but this is not the case. Tajbaksh et. al. [32] concludes that fine-tuning through transfer learning is better than training from scratch for a particular architecture. On the other hand, our proposition is to use custom architectures for different modalities. For example, LightOCT can be used for transfer learning on other OCT datasets, while a different architecture might be better suited for histopathology.

## 5. Data availability

LightOCT's pretrained networks for all the three datasets will be released online on the webpage of Dilip K. Prasad at the link: https://sites.google.com/site/dilipprasad/Source-codes.

## 6. Funding, *author contributions*, and disclosures

### *7.1 Funding*

### *7.2 Author contributions*

BSA, DKP, DSM, PS, and AS conceived the project and supervised this work. Experiments were performed by AB, AA, and VD. Sample was prepared by DQ. DKP and AB designed and developed the "LightOCT" architecture. Data were analyzed by AB and DKP. All the authors contributed towards the writing of the manuscript.

### *7.3 Disclosures*

Authors declare no competing interest.